\documentclass{article}

\usepackage{PRIMEarxiv}

\usepackage[utf8]{inputenc} 
\usepackage[T1]{fontenc}    
\usepackage{hyperref}       
\usepackage{url}            
\usepackage{booktabs}       
\usepackage{amsfonts}       
\usepackage{nicefrac}       
\usepackage{microtype}      
\usepackage{lipsum}
\usepackage{fancyhdr}       
\usepackage{graphicx} 
\usepackage{subcaption}
\graphicspath{{media/}} 
\usepackage{pdfrender}
\usepackage{amssymb}
\usepackage{amsbsy}

\title{Synthetic Data Generation Techniques for Developing AI-based Speech Assessments for Parkinson's Disease (A Comparative Study)}

\author{
  Mahboobeh Parsapoor (Mah Parsa) \\
\\
\\
}

\begin{document}
\maketitle

\begin{abstract}
Changes in speech and language are among the first signs of Parkinson's disease (PD). Thus, clinicians have tried to identify individuals with PD from their voices for years. Doctors can leverage AI-based speech assessments to spot PD thanks to advancements in artificial intelligence (AI). Such AI systems can be developed using machine learning classifiers that have been trained using individuals' voices. Although several studies have shown reasonable results in developing such AI systems, these systems would need more data samples to achieve promising performance. 
This paper explores using deep learning-based data generation techniques on the accuracy of machine learning classifiers that are the core of such systems.  
\end{abstract}


\section{Introduction}
Voice impairments are common problems for patients with Parkinson's disease (PD) \cite{Suppa2022} and other types of dementia \cite{parsapoor2023ai}. Many PD patients face difficulties in speech and language generation, particularly in modulating and articulating their voices or generating and inflecting verbs \cite{Liu2015, Altmann2011}. Accurately diagnosing language impairments associated with PD can significantly assist clinicians in identifying individuals with PD \cite{Liu2015, Altmann2011, Javeed2023}. Current advances in artificial intelligence (AI) \cite{Javeed2023, parsapoor2023ai, Parsapoor2023} have motivated clinicians to leverage AI-based speech assessments to detect speech assessments associated with PD and identify patients with PD. These AI should be developed based on machine learning  (ML) or deep learning classifiers, which are trained using extracted acoustic and linguistic features from patients' voices.  
Different studies have shown that traditional ML algorithms could reasonably classify acoustic features of patients with PD and healthy controls \cite{Suppa2022}. For example, in \cite{Suppa2022}, a Support Vector Machine (SVM) classifier, which was trained by acoustic features extracted using the Opensmile \cite{Eyben2010} software, could classify patients versus healthy controls with 76.9\% accuracy \cite{Suppa2022}. As another example, the authors of \cite{Nahar2021} demonstrated that training a Bagging algorithm with acoustic features selected using recursive feature elimination could classify vocal samples of patients and healthy controls with 82.35\% accuracy. In \cite{Alshammri2023}, the authors showed that using the set of acoustic features obtained combining Synthetic Minority Over-sampling Technique (SMOTE) for training MLP could lead to developing accurate classifiers which could distinguish patients from healthy controls with an overall accuracy of 98.31\%. 
In \cite{Pramanik2021}, the authors proposed an ML-based classifier to distinguish patients with PD from healthy controls with 94.10\%. 

Most of the above studies focused on using the available data samples or oversampling techniques to develop ML-based classifiers; however, using an appropriate data generation technique can increase the performance of ML-based classifiers. Thus, this study focuses on generating data samples using different methods and comparing their influences on ML performances.

\section{Methods}
The methods encompassed the following steps: (1)Utilizing data generation techniques; (2) Developing ML-based classifiers and explaining the effect of features on classifier's outputs.


\subsection{Datasets and data generation techniques}
\begin{figure*}
  \begin{minipage}[t]{.45\linewidth}
    \includegraphics[width=\linewidth]{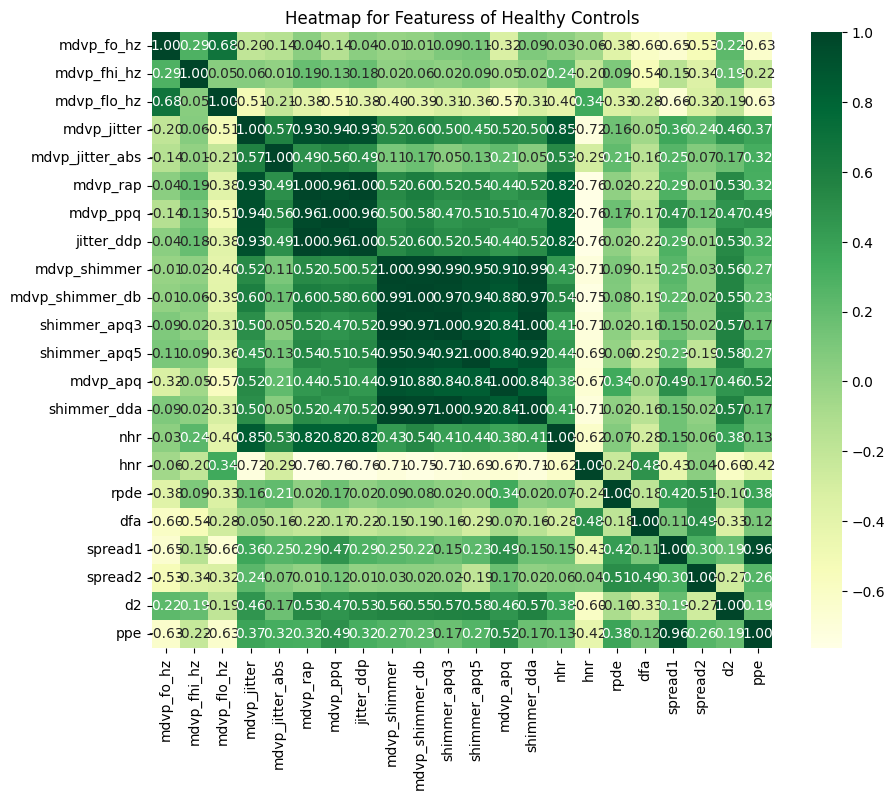}%
    \caption{Correlation heatmap of speech features of healthy individuals}%
    \label{fig:Healthy}
  \end{minipage}\hfil
  \begin{minipage}[t]{.45\linewidth}
    \includegraphics[width=\linewidth]{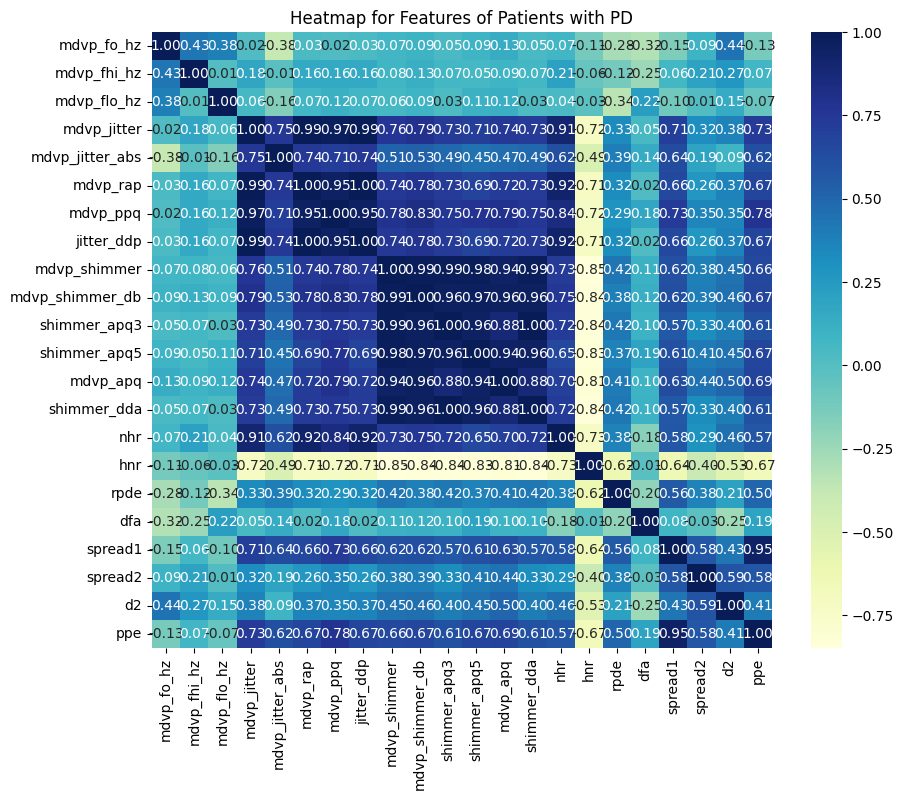}%
    \caption{Correlation heatmap of speech features of patients}%
    \label{fig:PD}
  \end{minipage}\hfil
\end{figure*}

The dataset of this paper is the benchmark dataset of acoustic features obtained from 195 vocal samples of 31 participants,  8 healthy control and 23 patients with Parkinson's disease (PD). Note that each participant has attended around six recordings
\cite{Little2009}. Figures \ref{fig:PD} and \ref{fig:Healthy} show the correlation heatmap of features for different subjects, healthy individuals and patients respectively. 

Each data sample consists of 22 acoustic features (see Table\ref{tab:Features} for description of each feature) that can be divided into (1) MDVP features, extracted using the Kay Pentax Multi-Dimensional Voice Program (MDVP); (2) Standard features; (3) Non-standard features including the correlation dimension (D2), the recurrence period density entropy (RPDE) and detrended fluctuation analysis (DFA); (4) New pitch-based feature named Pitch Period Entropy (PPE) \cite{Abiyev2016}.  

\begin{table}[h]
\centering
\caption{Description of 22 Speech Features}
\begin{tabular}{|p{0.30\linewidth}|p{0.65\linewidth}|}
\hline
 \textbf{Feature} & \textbf{Meaning}  \\
\hline
mdvp\_fo\_hz &  Average vocal fundamental frequency \\
mdvp\_fhi\_hz & Maximum vocal fundamental frequency\\
mdvp\_flo\_hz & Minimum vocal fundamental frequency\\
mdvp\_jitter &  Jitter in percentage\\
mdvp\_jitter\_abs & Absolute jitter in ms\\
mdvp\_rap & Relative amplitude perturbation\\
mdvp\_ppq & Five-point period perturbation quotient \\
jitter\_ddp & Average absolute difference of differences between jitter cycles\\
mdvp\_shimmer &Local shimmer\\
mdvp\_shimmer\_db &Local shimmer in dB\\
shimmer\_apq3 & Three-point amplitude perturbation quotient\\
shimmer\_apq5 & Five-point amplitude perturbation quotient\\
mdvp\_apq & 11-point amplitude perturbation quotient \\
shimmer\_dda & Average absolute differences between the amplitudes of consecutive periods\\
nhr & Noise-to-harmonics ratio\\
hnr & Harmonics-to-noise ratio\\
rpde & Recurrence period density entropy measure\\
dfa & Signal fractal scaling exponent of detrended fluctuation analysis\\
spread1 & Two nonlinear measures of fundamental\\
spread2 &Frequency variation\\
d2 & Correlation dimension\\
ppe & Pitch period entropy\\
\hline
\end{tabular}
\label{tab:Features}
\end{table}

\begin{table}[h]
\centering
\caption{List of features}

\begin{tabular}
{|p{0.30\linewidth}|p{0.2\linewidth}|p{0.2\linewidth}|p{0.2\linewidth}|}
\hline
 \textbf{Feature} & \textbf{TVAE} & \textbf{CTGAN} & \textbf{CopulaGAN} \\
\hline
mdvp\_fo\_hz & \pmb{\checkmark} & \pmb{\checkmark}& \pmb{\checkmark}\\
mdvp\_fhi\_hz & & \pmb{\checkmark}& \pmb{\checkmark}\\
mdvp\_flo\_hz &\pmb{\checkmark} & \pmb{\checkmark}& \\
mdvp\_jitter & & \pmb{\checkmark}& \\
mdvp\_jitter\_abs & &\pmb{\checkmark} & \\
mdvp\_rap &\pmb{\checkmark} & \pmb{\checkmark}& \pmb{\checkmark}\\
mdvp\_ppq & \pmb{\checkmark}& \pmb{\checkmark}& \pmb{\checkmark}\\
jitter\_ddp & \pmb{\checkmark}&\pmb{\checkmark} & \pmb{\checkmark}\\
mdvp\_shimmer &\pmb{\checkmark} & \pmb{\checkmark}& \pmb{\checkmark}\\
mdvp\_shimmer\_db &\pmb{\checkmark} & \pmb{\checkmark}& \\
shimmer\_apq3 & & \pmb{\checkmark}& \\
shimmer\_apq5 & \pmb{\checkmark}& \pmb{\checkmark}& \pmb{\checkmark}\\
mdvp\_apq & \pmb{\checkmark}&  \pmb{\checkmark}&\pmb{\checkmark}\\
shimmer\_dda & \pmb{\checkmark}& \pmb{\checkmark}& \\
nhr & \pmb{\checkmark}& \pmb{\checkmark}& \\
hnr & \pmb{\checkmark}& \pmb{\checkmark}& \\
rpde & \pmb{\checkmark}& \pmb{\checkmark}& \pmb{\checkmark}\\
dfa & & \pmb{\checkmark}& \\
spread1 & \pmb{\checkmark}& \pmb{\checkmark}& \pmb{\checkmark}\\
spread2 &\pmb{\checkmark} & \pmb{\checkmark}& \pmb{\checkmark}\\
d2 &\pmb{\checkmark} & \pmb{\checkmark}& \\
ppe & \pmb{\checkmark}& \pmb{\checkmark}& \pmb{\checkmark}\\
\hline
\end{tabular}
\label{tab:FeaturesList}
\end{table}

We employed different DL-based synthetic data generation techniques, including those proposed in \cite{tvae}. In more detail, we used the VAE-based Deep Learning data generation techniques, which have been referred to as TVAE, CTGAN, or conditional generator \cite{tvae} and CopulaGAN, which is a hybrid of the Gaussian Copula and the CTGAN algorithms, for generating data samples. Table\ref{tab:data} lists the total samples for the healthy and PD classes and the quality scores of different methods. Note that the quality score evaluates how well synthetic data captures mathematical properties in original data and indicates that the TVAE method can generate high-quality data compared to the other two methods.

\begin{table}[h]
\centering
\caption{Description of Generated Data Samples }
\begin{tabular}{|p{0.20\linewidth}|p{0.30\linewidth}|p{0.20\linewidth}|p{0.20\linewidth}|}
\hline
 \textbf{Methods} & \textbf{No. of Samples for Healthy Class} &\textbf{No. of Samples for Patients Class} & \textbf{Quality Score}\\
\hline
TVAE&92&80&82.43\%\\
CTGAN&86&86&74.48\%\\
CopulaGAN&78&94& 68.92\%\\
\hline
\end{tabular}
\label{tab:data}
\end{table}

\subsection{Developing ML-based classifiers}
Based on recommendations in other studies \cite{Little2009, Javeed2023}, we considered several traditional ML algorithms, including Random Forest (RF),  
Gradient Boosting (GB), Extreme Gradient Boosting (XGB), Adaptive Boosting (AdaBoost), 
Decision Tree (DT), Extremely Randomized Trees or Extra Trees (ET) and Support Vector Machine (SVM). 

\begin{figure*}
  \begin{minipage}[t]{.3\linewidth}
    \includegraphics[width=\linewidth]{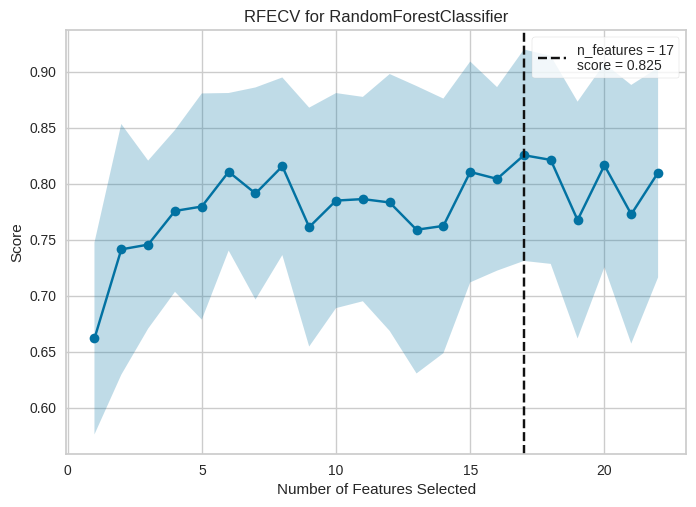}%
    \caption{The f1 score of the RF algorithm for different sets of features. Training and validating the algorithm using 22 features and 5-fold cross-validation}%
    \label{fig:FS_TVAE}
  \end{minipage}\hfil
  \begin{minipage}[t]{.3\linewidth}
    \includegraphics[width=\linewidth]{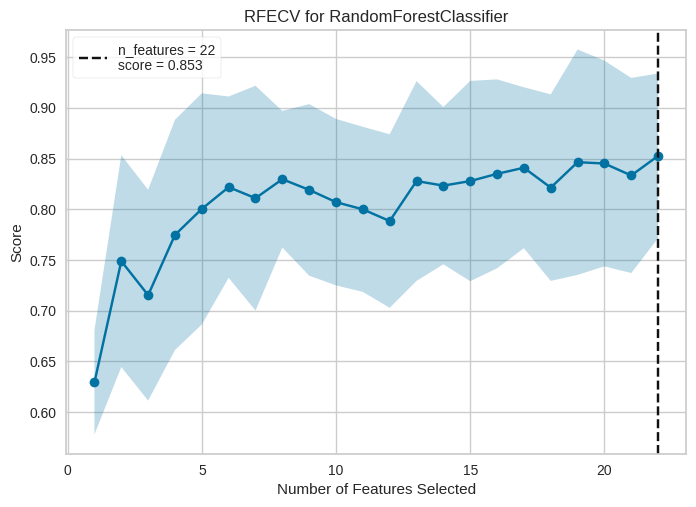}%
    \caption{The f1 score of the RF algorithm for different sets of features. Training and validating the algorithm using 22 and 8-fold cross-validation}%
    \label{fig:FS-CTGAN}
  \end{minipage}\hfil
  \begin{minipage}[t]{.3\linewidth}
    \includegraphics[width=\linewidth]{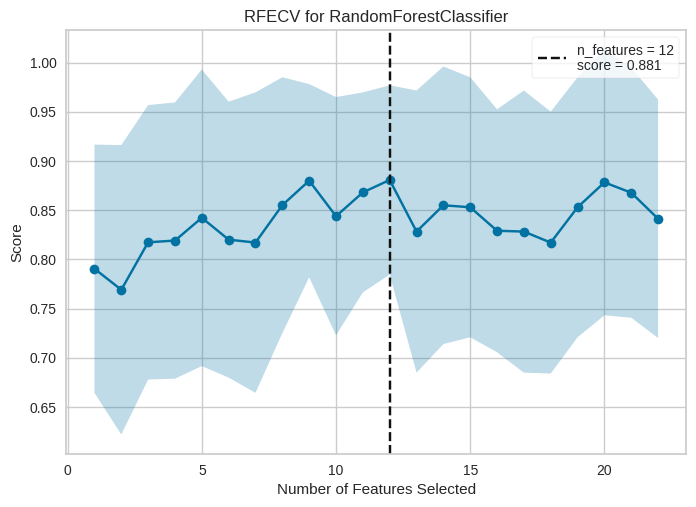}%
    \caption{The f1 score of the RF algorithm for different sets of features. Training and validating the algorithm using 22 features and 9-fold cross-validation}%
    \label{fig:Cap}
  \end{minipage}%
\end{figure*}

We employed recursive feature elimination (RFE), which \ has suggested cite{Cai2018, Remeseiro2019}, to select a subset of the most relevant features for the classification task that is classifying the speech features associated with patients from those that are related to healthy controls. RFE uses an ML algorithm (e.g., an SVM-based algorithm or an RF-based algorithm) as the learning algorithm and selects the optimal subset of features (i.e., the subset of features that are least essential features to be eliminated) based on the performance of the algorithm\cite{Jeon2020}. In our study, we considered an RF-based algorithm as the basis of our feature selection method, RFE and trained-validated the RF algorithm using 2 to 9 cross-fold validations. 
Figures \ref{fig:FS_TVAE}, \ref{fig:FS-CTGAN} and \ref{fig:Cap} describe the f1 score values from different sets of features for the datasets generated using the TVAE, CTGAN and CopulaGAN methods, respectively. We selected the number of cross-fold validations for each model that led to the most accurate classification results. The best f1 scores for the classifiers developed by the TVAE, CTGAN and CopulaGAN datasets using 5, 8 and 9 folds are around 0.83, 0.85 and 0.88, respectively. The classifiers achieved the accuracy mentioned above training with 17, 22 and 12 features, respectively.


\section{Results}
Tables \ref{table:resultsTVAE}, \ref{table:resultsCTGAN}, and \ref{table:resultsCopulaGAN} list the performance of various algorithms trained using generated datasets via different data generation techniques. The results indicate that, among other algorithms, training the ExtraTrees algorithm with data obtained from using the CopulaGAN technique can lead to the development of a high-performance classifier. Furthermore, the classifier has been trained with fewer features that might be more explainable than classifiers that could be developed using more features. Since studies \cite{AkbarzadehT2021, Mollaei2016} have indicated that the influence of PD on "Shimmer" related features and the reduction in both pitch height and range, the ML-base vocal system that considers these features to identify PD-related speech impairments could be more interpretable for clinicians.


\begin{table}[!ht]
\centering
\caption{Classifier Performance on validation set of the dataset generated using TVAE}
 \begin{tabular}
 {|p{0.22\linewidth}|p{0.16\linewidth}|p{0.16\linewidth}|p{0.16\linewidth}|p{0.16\linewidth}|}
  \hline
    \textbf{Classifier} & \textbf{Mean Accuracy} & \textbf{Mean F1-score} & \textbf{Std. Dev. Accuracy} & \textbf{Std. Dev. F1-score} \\
    \hline
    RandomForest & 0.797 & 0.795 & 0.069 & 0.070 \\
    GradientBoosting & 0.791 & 0.788 & 0.066 & 0.068 \\
    ExtraTrees & 0.838 & 0.838 & 0.052 & 0.051 \\
    AdaBoost & 0.808 & 0.806 & 0.072 & 0.074 \\
    DecisionTree & 0.791 & 0.789 & 0.043 & 0.045 \\
    XGB & 0.785 & 0.784 & 0.048 & 0.048 \\
    SVC & 0.814 & 0.813 & 0.046 & 0.046 \\
    \bottomrule
  \end{tabular}
\label{table:resultsTVAE}
\end{table}
\begin{table}[!ht]
\centering
\caption{Classifier performance validation set of the dataset generated using CTGAN}
\begin{tabular}
{|p{0.22\linewidth}|p{0.16\linewidth}|p{0.16\linewidth}|p{0.16\linewidth}|p{0.16\linewidth}|}
    \toprule
    \textbf{Classifier} & \textbf{Mean Accuracy} & \textbf{Mean F1-score} & \textbf{Std. Dev. Accuracy} & \textbf{Std. Dev. F1-score} \\
 \hline
    RandomForest & 0.873 & 0.872 & 0.059 & 0.060 \\
    GradientBoosting & 0.808 & 0.804 & 0.044 & 0.049 \\
    ExtraTrees & 0.873 & 0.872 & 0.081 & 0.081 \\
    AdaBoost & 0.832 & 0.831 & 0.078 & 0.080 \\
    DecisionTree & 0.780 & 0.776 & 0.085 & 0.088 \\
    XGB & 0.861 & 0.860 & 0.082 & 0.083 \\
    SVC & 0.855 & 0.852 & 0.060 & 0.061 \\
    \bottomrule
  \end{tabular}
\label{table:resultsCTGAN}
\end{table}

\begin{table}[!ht]
\centering
\caption{Classifier performance validation set of the dataset generated using the CopulaGAN method}
\begin{tabular}
{|p{0.22\linewidth}|p{0.16\linewidth}|p{0.16\linewidth}|p{0.16\linewidth}|p{0.16\linewidth}|}
\hline
\textbf{Classifier} & \textbf{Mean Accuracy} & \textbf{Mean F1-score} & \textbf{Std Accuracy} & \textbf{Std F1-score} \\
\hline
RandomForest & 0.906 & 0.905 & 0.078 & 0.079 \\
GradientBoosting & 0.835 & 0.825 & 0.092 & 0.107 \\
ExtraTrees & 0.906 & 0.906 & 0.078 & 0.078 \\
AdaBoost & 0.858 & 0.855 & 0.088 & 0.090 \\
DecisionTree & 0.811 & 0.802 & 0.102 & 0.113 \\
XGB & 0.883 & 0.881 & 0.076 & 0.077 \\
SVC & 0.895 & 0.894 & 0.071 & 0.072 \\
\hline
\end{tabular}
\label{table:resultsCopulaGAN}
\end{table}

\section{Conclusion}
This paper showed that by using the CopulaGAN techniques for generating data, we could develop classifiers with reasonable accuracy and explainability. 
The results we have shown in this paper are laboriously affected by (1) the heterogeneous characteristics of PD and (2) the restriction of accessing speech data of PD patients and healthy subjects, which has led to the unequal distribution of samples. Although we utilized undersampling techniques to address the limitations, increasing the number of actual speech samples would be essential to developing more accurate and robust classifiers. Furthermore, because age, gender and race might influence speech features, we would need a set of mechanisms to ensure that obtained decisions are fair.    

The future direction of this study encompasses two folds: (1) Developing an accurate, robust classifier adding more speech samples, using augmentation techniques including Mask Token Replacement (MTR) \cite{Madukwe2022}, utilizing few-shot ML algorithms, emotional learning-inspired ensemble classifier (ELiEC) \cite{6643988}
 and transfer learning techniques along with deep learning algorithms. (2) Deploying fair classifiers employing fairness techniques. 

\bibliographystyle{unsrt}  
\bibliography{references}

\end{document}